\documentclass[twoside]{article}
\usepackage{fleqn,espcrc2}

\newcommand{\AmS}{{\protect\the\textfont2
  A\kern-.1667em\lower.5ex\hbox{M}\kern-.125emS}}

\title{Covariant $\kappa-$Symmetry Gauge Fixing and the Classical Relation
Between
Physical Variables of the NSR String and the Type II GS
Superstring}

\author{Dmitriy V. Uvarov\address{NSC Kharkov Institute of Physics and
Technology \\
        61108 Kharkov, Ukraine}
        \thanks{E-mail uvarov@kipt.kharkov.ua,
d\_uvarov@hotmail.com}}

\begin{document}
\setcounter{page}{120}
\begin{abstract}
The main goal of this paper is the manifestly covariant
derivation of the classical relation between the gauge-fixed
Grassmann variables of the NSR string and the Type II GS
superstring. To this end we analyze the superembedding equation
for the Type II superstring to derive the relation between the
original variables of the NSR string and the Type II GS
superstring and, further, by means of  Lorentz harmonic variables
we fix $\kappa-$symmetry of the GS superstring in the manifestly
Lorentz covariant way.
\vspace{1pc}
\end{abstract}

\maketitle

\section{INTRODUCTION}

There exist two ways of unification of supersymmetry and string
theory. One is able either to introduce local supersymmetry on
the string worldsheet resulting in the Neveu-Schwarz-Ramond (NSR)
model \cite{NSR}, which can be regarded as the coupling of $2d$
induced supergravity with matter, or global supersymmetry in a
target-space which yields the Green-Schwarz (GS) superstring
\cite{GS}. The introduction of local supersymmetry in the
target-space leads to a superstring model interacting with
target-space supergravity. Models with both worldsheet (local)
and target-space (global or local) supersymmetry, the so called
spinning superstrings \cite{Gates}, \cite {spin}, in some sense
can be regarded as a product of NSR and GS models. They possess
wider spectrum of quantum states in comparison with the NSR and GS
models.

The NSR string and the GS superstring are in fact different string
theories at the classical level. It is remarkable, however, that
upon first quantization they describe the same set of quantum
states provided the NSR string is subject to GSO projection
\cite{GSO}, \cite{GSW}. Therefore it seems interesting to
establish a direct classical  relation between these models at
the level of variables and actions and to find a classical
analogue of the GSO projection. This problem was put forward  in
late 80-s by D.V. Volkov and his collaborators and the solution
was found for the case of superparticle and spinning particle
models \cite{VZ}, \cite{STV}, \cite{STVZ}. Since then interesting
results on the relation of the NSR and the GS superstring have
been obtained in \cite{Berk}, \cite{APT} and \cite{DP} which we
shall generalize in this contribution.

In the present paper following the guidelines of pioneer works
\cite{VZ}, \cite{STV}, \cite{STVZ} and using the methods that
were developed afterwards (the superembedding approach (see review
\cite{DS} and references therein) and Lorentz harmonic variables
\cite{Sokatchev} adapted to the description of strings
\cite{BZstring}, \cite{BZbrane}) we deduce the relation between
the physical variables (i.e. those surviving local  symmetry gauge
fixing) of the NSR string and the Type II GS superstring.  This requires
analysis of the superembedding equation for the $n=(1|1)$
worldsheet superspace embedded into the flat $D=10$ Type II
target-superspace to obtain the relation between the original
Grassmann variables of the models which include not only physical
but also pure gauge ones (Section 2). $SO(1,9)$ Lorentz covariant
$\kappa-$symmetry gauge fixing of the twistor-like Lorentz harmonic
formulation of the Type II GS superstring
\cite{BZstring} is the subject of Section 3. Finally in Section 4
we compile all these data to find the covariant relation between
the physical variables of the NSR string and the Type II GS
superstring.

\section{ANALYSIS OF THE SUPEREMBEDDING EQUATION FOR $n=(1|1)$
WORLDSHEET SUPERSPACE EMBEDDED INTO $D=10$ TYPE II TARGET-SUPERSPACE}

Let us consider the embedding of $n=(1|1)$ superworldsheet into flat
$D=10$ Type II target-superspace. $n=(1|1)$ worldsheet superspace is locally
parametrized  by the coordinates $z^M=(\xi^m,\eta^\mu)$, where
Grassmann coordinate $\eta^\mu$ is the $2d$ Majorana spinor that
consists of two one--component Majorana-Weyl spinors of opposite
chiralities (that explains the notation $n=(1|1)$).  The choice
of $n=(1|1)$ worldsheet superspace is dictated by the fact that
in such worldsheet superspace the NSR string is formulated.
$D=10$ Type II target-superspace is parametrized by coordinates
$Z^{\underline M}=(X^{\underline m},\theta^{1\underline\mu},
\theta^{2}_{\underline\mu})$ in the Type IIA case and by
$Z^{\underline M}=(X^{\underline m},\theta^{1\underline\mu},
\theta^{2\underline\mu})$ in the Type IIB case. From the
superembedding approach it follows that in order to describe
conventional string (brane) models one should impose certain
restrictions on the embedding, namely, the superembedding
equation \cite{DS} which reads that the pull-back of the
target-space supervielbein bosonic components along the
superworldsheet Grassmann directions vanishes
\begin{equation}\label{embed}
{D}_{a}Z^{\underline M}E_{\underline M}{}^{\underline a}(Z(z))=0,
\end{equation}
where $E_{\underline M}{}^{\underline A}=
(E_{\underline M}{}^{\underline a},E^1_{\underline M}{}^{\underline\alpha},
E^2_{\underline M}{}^{\underline\alpha} (E^2_{\underline
M\underline\alpha}))$
and $e_M{}^A=(e_M{}^a, e_M{}^\alpha)$
  are the target-space supervielbein and the
worldsheet superzweinbein respectively.
Their inverse are $E_{\underline B}{}^{\underline M}=
(E_{\underline b}{}^{\underline M}, E^1_{\underline \beta}{}^{\underline M},
E^2_{\underline \beta}{}^{\underline M}\ (E^{2\underline\beta\underline
M}))$,
$E_{\underline B}{}^{\underline M}E_{\underline M}{}^{\underline A}=
\delta_{\underline B}{}^{\underline A}$ and $e_B{}^M=(e_b{}^M,
e_\beta{}^M)$,
$e_B{}^Me_M{}^A=\delta_B{}^A$.
So that the worldsheet Grassmann covariant derivative entering (\ref{embed})
looks like $D_a=e_a{}^M\partial_M$. Using the explicit expression
for the bosonic components of the flat target-space supervielbein
\begin{eqnarray}
(IIA)\qquad\Pi^{\underline a}\equiv dZ^{\underline
M}E_{\underline M}{}^{\underline a}=\delta_{\underline
m}{}^{\underline a}\left(dX^{\underline m}\right.
\nonumber\\[0.3cm]\left.
- id\Theta^{1\underline\alpha}\sigma^{\underline
m}_{\underline\alpha\underline\beta}
\Theta^{1\underline\beta}-
id\Theta^{2}_{\underline\alpha}\tilde\sigma^{\underline
m\underline\alpha\underline\beta}
\Theta^{2}_{\underline\beta}\right),\\[0.3cm]
(IIB)\qquad\Pi^{\underline a}\equiv dZ^{\underline
M}E_{\underline M}{}^{\underline a}=\delta_{\underline
m}{}^{\underline a}\left(dX^{\underline m}\right.
\nonumber\\[0.3cm]
\left.- id\Theta^{1\underline\alpha}\sigma^{\underline
m}_{\underline\alpha\underline\beta}
\Theta^{1\underline\beta}
-id\Theta^{2\underline\alpha}\sigma^{\underline
m}_{\underline\alpha\underline\beta}
\Theta^{2\underline\beta}\right)
\end{eqnarray}
and splitting the tangent space indices of the superworldsheet in
light-like basis $e_M{}^A=(e_M{}^{\pm2},e_M{}^{\pm})$,
$e_A{}^M=(e_{\pm2}{}^M, e_{\pm}{}^M)$
one obtains from (\ref{embed})
\begin{eqnarray}
\lefteqn{D_{\pm}X^{\underline m}}
\nonumber\\[0.3cm]
\lefteqn{-iD_{\pm}\Theta^{1\underline\alpha}\sigma^{\underline
m}_{\underline\alpha\underline\beta}\Theta^{1\underline\beta}-iD_{\pm}\Theta^{2}_{\underline\alpha}\tilde\sigma^{\underline
m\underline\alpha\underline\beta}\Theta^{2}_{\underline\beta}=0}
\end{eqnarray}
for the Type IIA case and
\begin{eqnarray}
\lefteqn{D_{\pm}X^{\underline m}}\nonumber\\[0.3cm]
\lefteqn{-iD_{\pm}\Theta^{1\underline\alpha}\sigma^{\underline
m}_{\underline\alpha\underline\beta}\Theta^{1\underline\beta}-iD_{\pm}\Theta^{2\underline\alpha}\sigma^{\underline
m}_{\underline\alpha\underline\beta}\Theta^{2\underline\beta}=0}
\end{eqnarray}
for the Type IIB case. $n=(1|1)$ supergravity in $2d$ is well
studied
\cite{Howe} and the basic fact is that it is
superconformally flat. Thus, the component expansion of the
worldsheet superfields $X^{\underline m}(z^M)$,
$\Theta^{1,2\underline\alpha}(z^M)$ acquire the most simple form
\begin{eqnarray}
X^{\underline m}(\xi^m,\eta^\mu)=x^{\underline m}(\xi^m)+
\frac{i}{\sqrt{8}}\eta^+\psi^{\underline m}_+(\xi^m)\nonumber\\
+\frac{i}{\sqrt{8}}\eta^-\psi^{\underline
m}_-(\xi^m)+i\eta^+\eta^-F^{\underline
m}(\xi^m),\qquad\quad
\end{eqnarray}
\begin{eqnarray}
\Theta^{1,2\underline\alpha}(\xi^m,\eta^\mu)=
\theta^{1,2\underline\alpha}(\xi^m)+\eta^+\lambda^{1,2\underline\alpha}_+(\xi^m)\nonumber\\
[0.2cm]
+\eta^-\lambda^{1,2\underline\alpha}_-(\xi^m)+i\eta^+\eta^-\rho^{1,2\underline\alpha}(\xi^m).
\qquad\qquad
\end{eqnarray}
$x^{\underline m}$ and $\theta^{1,2\underline\alpha}$
are the ordinary GS variables, $\psi^{\underline m}_{\pm}$ are
the NSR Grassmann variables,
$\lambda^{1,2\underline\alpha}_{\pm}$ are the stringy twistor-like
variables, and
$F^{\underline m}$ and $\rho^{1,2\underline\alpha}$ are auxiliary fields.
For further analysis we impose the chirality conditions on
$\Theta^{1,2\underline\alpha}$ superfields \cite{VZ}
\begin{equation}\label{ms}
{D}_-\Theta^{1\underline\alpha}={D}_+\Theta^{2\underline
\alpha}=0,
\end{equation}
which on the  component level are equivalent to
\begin{equation}\label{redund}
\lambda^{1\underline\alpha}_-=\rho^{1\underline\alpha}=0,\ \
\lambda^{2\underline\alpha}_+=\rho^{2\underline\alpha}=0;
\end{equation}
\begin{eqnarray}\label{chir1}
\partial_{-2}\theta^{1\underline\alpha}=
\partial_{-2}\lambda^{1\underline\alpha}_+\equiv\partial_{-2}\lambda^{\underline\alpha}_+=0,\\[0.3cm]
\label{chir2}\partial_{+2}\theta^{2\underline\alpha}
=\partial_{+2}\lambda^{2\underline\alpha}_-\equiv\partial_{+2}\lambda^{\underline\alpha}_-=0.
\end{eqnarray}
The conditions (\ref{ms}) contain equations of motion and thus put
the variables on the mass shell. When eq. (\ref{redund}) is
imposed the superembedding equation yields
\begin{equation}\label{twistor}
\psi^{\underline m}_-=
\sqrt{8}\lambda^{\underline\alpha} _-\sigma^{\underline m}_{\underline\alpha
\underline\beta}
\theta^{2\underline\beta},\ \
\psi^{\underline m}_+=
\sqrt{8}\lambda^{\underline\alpha}_+\sigma^{\underline m}_{\underline\alpha
\underline\beta}
\theta^{1\underline\beta};
\end{equation}
\begin{equation}\label{Vir}
\Pi^{\underline m}_{\pm2}=\lambda_{\pm}\sigma^{\underline m}\lambda_{\pm};
\end{equation}
\begin{equation}
F^{\underline m}=0.
\end{equation}
For the Type IIA case we have $\psi^{\underline
m}_-=\sqrt{8}\lambda_{\underline\alpha-}\tilde\sigma^{\underline
m\underline\alpha \underline\beta}\theta^{2}_{\underline\beta}$,
$\Pi_{-2}^{\underline m}=\lambda_{\underline\alpha
-}\tilde\sigma^{\underline
m\underline{\alpha\beta}}\lambda_{\underline\beta -}$. Applying
then equations (\ref{chir1}) and (\ref{chir2}) one gets the NSR
string fermionic equations of motion
\begin{equation}\label{NSReqm}
\partial_{+2}\psi^{\underline m}_-=0,\ \partial_{-2}\psi^{\underline m}_+=0.
\end{equation}

Let us outline some properties of the obtained formulae.
Representation (\ref{twistor}) connects in a natural way  the
Grassmann variables of the NSR string and the Type II GS
superstrings. Its main drawback is the mismatch between the
number of degrees of freedom on the  l.h.s and the r.h.s. The
balance is, however,  restored on the constraint shell. Indeed,
the NSR string Grassmann vectors
$\psi^{\underline m}_{\pm}$ contain $9+9$ components as a result
of two supercurrent constraints. However, as will be seen below,
a proper solution of these constraints ensures dropping away the
two extra components of $\psi^{\underline m}_{\pm}$ from the NSR
string action. So, only  $8+8$ physical components of
$\psi^{\underline m}_{\pm}$ contribute to the action. On the other
hand, among $16+16$ components of the two $10D$ MW spinors
$\theta^{1,2\underline\alpha}$ there remain $8+8$ components after explicit
gauge
fixing $\kappa-$symmetry (see Sec.3). Thus, on the constraint
shell there is the same number of the Grassmann degrees of
freedom in  both formulations of string theory, as it should be.
In  Section  4 we will find manifest expressions for the physical
variables in the NSR and GS model and relate them to  each other.
Representation (\ref{Vir}) solves the Type II GS superstring
Virasoro constraints since the vectors
$\lambda_+\sigma^{\underline m}\lambda_+$ and
$\lambda_-\sigma^{\underline m}\lambda_-$ are light-like due to
the famous $10D$ permutation relation
\begin{equation}\label{perm}
\sigma^{\underline m}_{\underline\alpha\underline\beta}
\sigma_{\underline m\underline\gamma\underline\delta}+
\sigma^{\underline m}_{\underline\alpha\underline\delta}
\sigma_{\underline m\underline\beta\underline\gamma}+
\sigma^{\underline m}_{\underline\alpha\underline\gamma}
\sigma_{\underline m\underline\delta\underline\beta}=0.
\end{equation}
The NSR string and  GS superstring equations of motion are satisfied by
virtue
of (\ref{chir1}-\ref{Vir}).

\section{TWISTOR-LIKE LORENTZ HARMONIC FORMULATION OF TYPE II GS
SUPERSTRINGS AND COVARIANT $\kappa-$SYMMETRY FIXING}

The twistor-like Lorentz harmonic formulation
of the Type IIB GS superstring, which is classically equivalent to
the original formulation was constructed in \cite{BZstring}:
\begin{eqnarray}
\lefteqn{{\displaystyle S=\int
e\left(-(\alpha^\prime)^{-1/2}e^{m}_au^{a}_{\underline m}
\Pi^{\underline m}_m+c\right)}}\nonumber\\[0.3cm]
\lefteqn{{\displaystyle-\frac{1}{c\alpha^\prime}\int
\epsilon^{mn}i\Pi^{\underline m}_m\left(\partial_n\theta^1\sigma_{\underline
m}\theta^1-\partial_n\theta^2\sigma_{\underline
m}\theta^2\right)}}\nonumber\\[0.3cm]
\lefteqn{{\displaystyle+\frac{1}{c\alpha^\prime}\int
\epsilon^{mn}\partial_m\theta^1\sigma_{\underline
m}\theta^1\partial_n\theta^2\sigma_{\underline m}\theta^2}.}\label{gs}
\end{eqnarray}
In the Type IIA case one should replace
$\partial_n\theta^{2\underline\alpha}\sigma_{\underline
m\underline\alpha\underline\beta }
\theta^{2\underline\beta}$ with $\partial_n\theta^2_{\underline\alpha}
\tilde \sigma_{\underline m}^{\underline\alpha\underline\beta
}\theta^2_{\underline\beta}$.
In addition to the variables present in the standard GS
superstring formulation \cite{GS} it contains the worldsheet
zweinbein $e^{m}_a$ and the light-like Lorentz frame vectors
$u^{a}_{\underline m}$ tangent to the string worldsheet. These
light-like Lorentz frame vectors together with
$u^{\underline m i}$ $(i=1,...,8)$,  orthogonal to
the worldsheet, constitute a complete orthonormal basis
\begin{eqnarray}
\lefteqn{u^{\pm2}\cdot u^{\pm2}=0, u^{+2}\cdot u^{-2}=2,}\nonumber\\
\lefteqn{u^{\pm2}\cdot u^{i}=0, u^{i}\cdot
u^{j}=-\delta^{ij}}\label{harmonics}
\end{eqnarray}
which one can use to expand any  $D=10$ Minkowski vector. The
Lorentz frame vectors can be presented as  bilinear combinations
of the spinor harmonics
$v^{(\underline\alpha)}_{\underline\alpha}=(v^{-}_{\underline\alpha\dot
A},v^{+}_{\underline\alpha A})$ or their inverse
$(v^{-1})^{\underline\alpha}_{(\underline\alpha)}=(v^{\underline\alpha-}_A,v^{\underline\alpha+}_{\dot
A})$ ($(v^{-1})^{\underline\alpha}_{(\underline\alpha)}
v^{(\underline\beta)}_{\underline\alpha}=\delta^{(\underline\beta)}_{(\underline\alpha)}$)
\cite{harmonics}, \cite{BZstring}, \cite{BZbrane}.
$SO(1,9)$ Lorentz group Cartan forms with indices decomposed in a
$SO(1,1)\times SO(8)$ manner are defined by the following relations
\begin{eqnarray}
\lefteqn{\Omega^{(\underline k)(\underline l)}=u^{\underline m(\underline
k)}du_{\underline m}^{(\underline l)}=\left(\Omega^{(0)}, \Omega^{\pm2i},
\Omega^{ij}\right):}\nonumber\\
\lefteqn{\Omega^{(0)}=-\frac12\Omega^{+2-2}=\frac12u^{\underline
m-2}du^{+2}_{\underline m}=-\frac12u^{\underline m+2}du^{-2}_{\underline
m},}\nonumber\\
\lefteqn{\Omega^{\pm2i}=u^{\underline m\pm2}du^i_{\underline m},}\nonumber\\
\lefteqn{\Omega^{ij}=u^{\underline mi}du^j_{\underline m}.}\label{cartan}
\end{eqnarray}
 From the embedding theory point of view
\cite{Eis},\cite{Barbashov} $\Omega^{ij}_{m}$ can be identified
with the torsion (third fundamental form) components,
$\Omega^{+2i}_{m}$ and $\Omega^{-2i}_{m}$ with the second
fundamental form components and $\Omega^{(0)}_{m}$ with the $2d$
spin connection. Integrability conditions of eqs. (\ref{cartan})
are the Gauss, Peterson-Kodacci and Ricci equations
\cite{Eis},\cite{Barbashov}.

The action (\ref{gs}) is invariant under $\kappa-$symmetry
transformations with local parameters $\kappa^+_A$ and
$\kappa^-_{\dot A}$ (see \cite{BZbrane}). For the  Type IIA case
$\kappa-$symmetry transformations read:
\begin{eqnarray}
\lefteqn{\delta\theta^{1\underline\alpha}=v^{\underline\alpha-}_A\kappa^+_A,\
\
\delta\theta_{\underline\alpha}^2=v^+_{\underline\alpha A}\kappa^-_{
A},}\nonumber\\[0.3cm]
\lefteqn{\delta x^{\underline m}=i\left(\kappa^+_A
v^{\underline\alpha-}_A\sigma^{\underline
m}_{\underline\alpha\underline\beta}
\theta^{1\underline\beta}+\kappa^-_{A}v^+_{\underline\alpha A}
\tilde\sigma^{\underline
m\underline\alpha\underline\beta}\theta_{\underline\beta}^2\right),}
\nonumber\\[0.3cm]
\lefteqn{\delta\left(ee^{m+2}\right)=
\frac{4i}{c(\alpha^\prime)^{1/2}}\kappa^+_A\varepsilon^{mn}\partial_n
\theta^{1\underline\alpha}v^{+}_{\underline\alpha A},}\nonumber\\[0.3cm]
\lefteqn{\delta\left(ee^{m-2}\right)
=-\frac{4i}{c(\alpha^\prime)^{1/2}}\kappa^-_{A}\varepsilon^{mn}\partial_n
\theta_{\underline\alpha}^2v^{\underline\alpha-}_{A},}\nonumber\\[0.3cm]
\lefteqn{\delta u^{+2}_{\underline m}
=-\frac{2i}{c(\alpha^\prime)^{1/2}}e^{m+2}w^{i}_m u^{i}_{\underline
m},}\nonumber\\[0.3cm]
\lefteqn{\delta u^{-2}_{\underline m}
=\frac{2i}{c(\alpha^\prime)^{1/2}}e^{m-2}w^{i}_m u^{i}_{\underline
m},}\label{kappaA}
\end{eqnarray}
where \\
$w^{i}_m=-\left(\kappa^+_B\gamma^i_{B\dot B}\partial_m
\theta^{1\underline\alpha}
v^{-}_{\underline\alpha\dot B}+\right.$$\left.
\kappa^-_{B}\gamma^i_{B\dot B}\partial_m
\theta_{\underline\alpha}^2v^{\underline\alpha+}_{\dot B}\right)$.

To gauge fix $\kappa-$symmetry it is  useful to expand the
Grassmann variables in the basis of the spinor harmonics:
\begin{eqnarray}
\lefteqn{\theta^{1\underline\alpha}=v^{\underline\alpha-}_A \theta^{1+}_A+
v^{\underline\alpha+}_{\dot A}\theta^{1-}_{\dot A},}\nonumber\\
\lefteqn{\theta^{2}_{\underline\alpha}=v_{\underline\alpha A}^+
\theta^{2-}_A+
v^{-}_{\underline\alpha\dot A}\theta^{2+}_{\dot A}}\label{newtetaA}
\end{eqnarray}
for the Type IIA case and
\begin{equation}\label{newtetaB}
\theta^{1(2)\underline\alpha}=v^{\underline\alpha-}_A \theta^{1(2)+}_A+
v^{\underline\alpha+}_{\dot A}\theta^{1(2)-}_{\dot A}
\end{equation}
for the Type IIB case.

Using (\ref{kappaA}) it is straightforward to show that among the
variables introduced in (\ref{newtetaA}) and (\ref{newtetaB})
$\theta^{1+}_A$,
$\theta^{2-}_{A}$ (IIA) and
$\theta^{1+}_A$, $\theta^{2-}_{\dot A}$ (IIB) are pure gauge,
thus it is admissible to impose the following covariant
$\kappa-$symmetry gauge fixing conditions:
\begin{equation}
\theta^{1+}_A=\theta^{2-}_{A}=0,\qquad\qquad\qquad\qquad(IIA)
\end{equation}
\begin{equation}
\theta^{1+}_A=\theta^{2-}_{\dot A}=0.\qquad\qquad\qquad\qquad(IIB)
\end{equation}

In this gauge the remaining variables
$\theta^{1-}_{\dot A}\equiv\theta^-_{\dot A}$, $\theta^{2+}_{\dot
A}\equiv\theta^+_{\dot A}$
(IIA) and $\theta^{1-}_{\dot A}\equiv\theta^-_{\dot A}$,
$\theta^{2+}_A\equiv\theta^+_A$ (IIB)
are $\kappa-$invariant. Note, that they are the worldsheet MW spinors,
whereas original variables $\theta^{\underline\alpha1,2}$ were the
worldsheet scalars.

In conclusion let us
present $\kappa-$symmetry-fixed version of (\ref{gs}). To this
end analogously to the Grassmann variables we expand the bosonic
coordinate $x^{\underline m}$ in the  basis of the vector
harmonics
\begin{equation}\label{x}
x^{\underline m}=\frac12 u^{\underline m+2}x^{-2}+
\frac12 u^{\underline m-2}x^{+2}-u^{i}_{\underline m}x^{i}.
\end{equation}
Then the derivatives $\partial_m x^{\underline m}$ entering the
action (\ref{gs}) transform into the covariant  derivatives
containing the
$SO(1,9)$ Cartan forms constructed from the harmonics
\begin{equation}
D_m x^{(\underline n)}=\partial_m x^{(\underline n)}+
\Omega_{m}^{(\underline n)(\underline l)} x_{(\underline l)}.
\end{equation}
The covariant derivatives of the fermionic variables are
\begin{eqnarray}
D_{m}\theta^+_A=\partial_{m}\theta^+_A-\frac12\Omega^{(0)}_m\theta^+_A-\frac14\Omega^{ij}_{m}
\gamma^{ij}_{AB}\theta^+_B,\\
D_{m}\theta^{\pm}_{\dot A}=\partial_{m}\theta^{\pm}_{\dot A}\mp
\frac12\Omega^{(0)}_m\theta^{\pm}_{\dot A}-\frac14\Omega^{ij}_{m}
\tilde\gamma^{ij}_{\dot A\dot B}\theta^{\pm}_{\dot B}.
\end{eqnarray}
Then the $\kappa-$symmetry gauge fixed action for the Type II GS
superstring takes the form
\begin{eqnarray}
\lefteqn{\displaystyle{S_{fixed}=}}\nonumber\\[0.3cm]
\lefteqn{{\displaystyle\int e\left[-(\!\alpha^\prime)^{-1/2}e^m_{-2}
\left(\!D_{m}x^{-2}\!-\!2iD_{m
}\theta^-\!\cdot\!\theta^-\!\right)\!\right.}}\nonumber\\[0.3cm]
\lefteqn{{\displaystyle\left.-\!(\!\alpha^\prime)^{-1/2}e^m_{+2}
\left(D_{m}x^{+2}\!-\!2iD_{m}\theta^+\!\cdot\!\theta^+\!\right)\!+\!c\right]}}\nonumber\\[0.3cm]
\lefteqn{\displaystyle{\!\!-\frac{i}{c\alpha^\prime}\int\!\!\epsilon^{mn}\!
\left[\!\left(\!D_{m}x^{+2}\!-\!2iD_{m}\theta^+\!\!\cdot\!\theta^+\!\right)\!D_{n}
\theta^-\!\!\cdot\!\theta^-\!\!\right.}}\nonumber\\[0.3cm]
\lefteqn{{\displaystyle\left.-\!\!\left(\!D_{m}x^{-2}\!-\!2i D_{m}
\theta^-\!\!\cdot\!\theta^-\!\right)\!D_{n}\theta^+\!\!\cdot\!\theta^+\!\!\right.}}\nonumber\\[0.3cm]
\lefteqn{{\displaystyle\left.-2i D_{m}\theta^-\!\!\cdot\!\theta^-\!D_{n}
\theta^+\!\!\cdot\!\theta^+\right.}}\nonumber\\[0.3cm]
\lefteqn{\displaystyle{\left.
-\frac12\!\left(\!D_{m}x^{i}\!\!-\!\!\frac14\sum\limits_{\pm}\!\left(\!\theta^{\pm}
\gamma^{ij}\theta^{\pm}\!\right)\!\Omega^{\mp2j}_{m}\!\right)\!\!\times\right.}}
\nonumber\\[0.3cm]
\lefteqn{{\displaystyle\left.\left(\!\left(\!\theta^-
\tilde\gamma^{ij}\theta^-\!\right)\!\Omega^{+2j}_{n}\!-\!\left(\!\theta^+\gamma^{ij}
\theta^+\!\right)\!\Omega^{-2j}_{n}\!\right)\!\right].}}\label{gsfixed}
\end{eqnarray}
The form of the action (\ref{gsfixed}) resembles that of Ref.
\cite{Sokatchev} for the superparticles. The action (\ref{gsfixed})
contains the following light-cone-like terms quadratic in
$\theta$
\begin{eqnarray}
\lefteqn{{\displaystyle S_{l.c.}\!=}}\nonumber\\[0.3cm]
\lefteqn{{\displaystyle\!\frac{2i}{(\alpha^\prime){}^{1/2}}\!\!\int\!\!
e\!\!\sum
\limits_{\pm}\!\left(\!1\!+\!\frac{1}{c(\alpha^\prime){}^{1/2}}\!D_{\pm2}x^{\pm2}\!\right)\!\!
D_{\mp2}\!\theta^{\mp}\!\!\!\cdot\!\theta^{\mp}}}\label{gslightcone}
\end{eqnarray}

\section{RELATION BETWEEN PHYSICAL VARIABLES OF THE NSR STRING AND TYPE II
GS SUPERSTRING}

First let us establish the connection between the Lorentz
harmonic variables and the commuting spinors
$\lambda^{\underline\alpha}_{\pm}$ of Section 2. The corresponding relation in
the case of
$n=(8|8)$ worldsheet superspace was established in \cite{BPSTV}.
Note that the variation of the action (\ref{gs}) with
respect to the zweinbeins and harmonics produces component
embedding equation
\begin{equation}\label{gsembed1}
\Pi^{\underline m}_m=\frac{c(\alpha^\prime){}^{1/2}}{2}
\left(e^{-2}_m u^{\underline m+2}+e^{+2}_m u^{\underline m-2}\right).
\end{equation}
Upon choosing the conformal gauge for the zweinbein
$e^a_m=e^{-\phi}\delta^a_m$ equation (\ref{gsembed1}) transforms into
\begin{equation}\label{gsembed}
\Pi^{\underline m}_{\pm2}=
{\textstyle\frac{c(\alpha\prime)^{1/2}}{2}}e^{-\phi}u^{\underline m\mp2}.
\end{equation}
Equation (\ref{gsembed}) coincides with (\ref{Vir}) only if
\begin{equation}\label{coin}
\lambda_{\pm}\sigma^{\underline m}\lambda_{\pm}=
\frac{c(\alpha^\prime){}^{1/2}}{2}e^{-\phi}u^{\underline m\mp2},
\end{equation}
which establishes the link with the discussion of the Section 3.
To analyze the consequences of (\ref{coin}) let us expand
$\lambda^{\underline\alpha}_{\pm}$ in the basis of the spinor harmonics
\begin{eqnarray}
\lefteqn{\lambda^{\underline\alpha} _+\!=\!\! k
\left(v^{\underline \alpha -}_A\lambda_A+v^{\underline \alpha+}_{\dot A}
\lambda_{+2\dot A}\right)\!,}\nonumber\\
\lefteqn{\lambda^{\underline\alpha} _-\!=\!\! k\left(v^{\underline\alpha
-}_A
\lambda_{-2A}+v^{\underline\alpha+}_{\dot A}\lambda_{\dot
A}\right)\!,}\label{newlambda}
\end{eqnarray}
where
$k=\left({\textstyle\frac{c(\alpha\prime)^{1/2}}{2}}\right)^{1/2}\!\!\!
e^{-\phi/2}$.
Then one finds that $\lambda_{+2\dot A}=\lambda_{-2A}=0$ and
$\lambda_A\lambda_A=1, \lambda_{\dot A}\lambda_{\dot A}=1$ for the Type IIB
case.
Corresponding relations for the Type  IIA case are
$\lambda_{\pm2\dot A}=0$ and
$\lambda^{1}_A\lambda^{1}_A=\lambda^{2}_A\lambda^{2}_A=1$.

The connection between  $\kappa-$symmetry gauge fixed GS variables
and NSR variables can be established upon the substitution of
\begin{eqnarray}
\lefteqn{\theta^{\underline\alpha1}=v^{\underline\alpha +}_{\dot
A}\theta^-_{\dot A},\ \theta^{\underline\alpha 2}=v^{\underline\alpha
-}_A\theta^+_A,}\nonumber\\
\lefteqn{\lambda^{\underline\alpha}_+=k v^{\underline\alpha -}_A\lambda_A,\
\lambda^{\underline\alpha}_-=k v^{\underline\alpha +}_{\dot A}\lambda_{\dot
A}}\label{rep}
\end{eqnarray}
into (\ref{twistor}) and the expansion of $\psi^m_{\pm}$ in the
basis of the vector harmonics
\begin{equation}\label{psi}
\psi^{\underline m}_{\pm}=
\frac12 u^{m+2}\varphi^{-2}_{\pm}+\frac12
u^{m-2}\varphi^{+2}_{\pm}-u^{i}_m\varphi^{i}_{\pm}.
\end{equation}
As a result we obtain
\begin{eqnarray}
\lefteqn{\varphi^{+2}_\pm=\varphi^{-2}_\pm=0,}\nonumber\\
\lefteqn{\varphi^i_+=\!\!-\sqrt{8}k\lambda_A\gamma ^i_{A\dot
B}\theta^-_{\dot B},\varphi^i_-=\!\!-\sqrt{8}k\lambda_{\dot A}\tilde \gamma
^i_{\dot A B}\theta^+_B.}\label{fix}
\end{eqnarray}
For the Type IIA case we have:
\begin{eqnarray}
\lefteqn{\theta^{\underline\alpha 1}=v^{\underline\alpha +}_{\dot
A}\theta^-_{\dot A}, \theta_{\underline\alpha}^2=v^-_{\underline\alpha\dot
A}\theta^+_{\dot A},}\nonumber\\
\lefteqn{\lambda^{\underline\alpha}_+=k v^{\underline\alpha -}_A\lambda^1_A,
\lambda_{\underline\alpha -}=k v^+_{\underline\alpha A}\lambda^2_{A},}
\end{eqnarray}
so
\begin{eqnarray}
\lefteqn{\varphi^{+2}_\pm=\varphi^{-2}_\pm=0,}\nonumber\\
\lefteqn{\varphi^i_+=\!\!-\sqrt{8}k\lambda^1_A\gamma ^i_{A\dot
B}\theta^-_{\dot B},\varphi^i_-=\!\! -\sqrt{8}k\lambda^2_{A}\gamma^i_{A\dot
B}\theta^+_{\dot B}.}
\end{eqnarray}

\section{CONCLUSION}
We have considered the superembedding equation for $n=(1|1)$
worldsheet superspace embedded into flat $D=10$ Type II
target-superspace. It was shown to contain the relation
(\ref{twistor}) between the original variables of the NSR string
and the Type II GS superstring, as well as the solution
(\ref{Vir})  to the Type II GS superstring Virasoro constraints.
Upon  Lorentz covariant gauge fixing
$\kappa-$symmetry with the use of the twistor-like Lorentz harmonic
variables, which amounts to covariantizing the light-cone gauge,
eq.(\ref{twistor}) reduces to the relation between
$\kappa-$symmetry gauge fixed variables $\theta^{\pm}_{\dot A}$
(Type IIA) and  $\theta^+_A, \theta^-_{\dot A}$ (Type IIB) of the
GS superstring, and the transverse physical variables
$\varphi^{i}_\pm$ of the NSR string.

\section{ACKNOWLEDGEMENTS}
The author would like to thank A.A. Zheltukhin for the numerous
valuable discussions, I.A.~Bandos and D.~Polyakov for the
interest to the work and stimulating discussions, and the Abdus
Salam ICTP, where the part of this work was done, for the warm
hospitality. The work was supported by Ukrainian State Foundation
for Fundamental Research.

\end{document}